\documentclass[modern]{aastex62}

\usepackage{amsmath}
\usepackage{graphicx}
\usepackage[capitalise]{cleveref}

\newcommand{\bleriot}{Bl{\'e}riot}
\newcommand{\rhill}{\text{$h$}}
\newcommand{\ak}{\text{$aK$}}
\begin{document} 

%\journalinfo{\textit{Astrophys.~J.~Lett.} \textbf{718}, L92--L96 (2010).}
%\submitted{}
\title{Analyzing \bleriot's propeller gaps in Cassini NAC images} 

\author[0000-0002-5864-5405]{Holger Hoffmann}
\affiliation{Institute of Physics and Astronomy, University of Potsdam, Potsdam, Germany.}
\author{Michael Seiler}
\affiliation{Institute of Physics and Astronomy, University of Potsdam, Potsdam, Germany.}
\author{Martin Sei{\ss}}
\affiliation{Institute of Physics and Astronomy, University of Potsdam, Potsdam, Germany.}
\author{Frank Spahn}
\affiliation{Institute of Physics and Astronomy, University of Potsdam, Potsdam, Germany.}

\correspondingauthor{Holger Hoffmann}
\email{hohoff@uni-potsdam.de}

\begin{abstract}
Among the great discoveries of the Cassini mission are the propeller-shaped
structures created by small moonlets embedded in Saturn's dense rings. We
analyze images of the sunlit side of Saturn's outer A ring, which show the 
propeller \bleriot{} with clearly visible partial propeller gaps. By determining
radial brightness profiles at different azimuthal locations, we obtain the evolution
of the gap minimum downstream of the moonlet. From the radial separation of the
partial propeller gaps we estimate the Hill radius of \bleriot{} to be about 400\,m.
Further, we fit the analytic solution from \citet{sremcevic2002mnras} describing
the azimuthal evolution of the surface mass density in the propeller gap region
to the azimuthal gap evolution obtained from Cassini images. From these fits,
we estimate a kinematic shear viscosity in the range of $60\,\text{cm}^2/\text{s}$
to $100\,\text{cm}^2/\text{s}$ in \bleriot's ring region. These values are consistent
with the parametrization given by \citet{daisaka2001icarus} and agree well with
values estimated for the Encke gap edge \citep{tajeddine2017apjs, graetz2018apj}.
% https://journals.aas.org/authors/keywords2013.html
\end{abstract}

\section{Introduction}

More than 150 propeller structures have been found in images taken by the
Cassini spacecraft \citep{tiscareno2006nature, sremcevic2007nature, tiscareno2008aj}.
Although most propeller structures have been identified in the propeller belts,
a radial region between 126,750 km and 132,000 km in the mid A ring, very large
propellers extending a few kilometers radially and up to several thousand
kilometers azimuthally were detected in the region between the Encke and Keeler
gaps \citep{tiscareno2010apjl}. The dimensions of these trans-Encke propellers
suggest subkilometer sized moonlets.

The ring-embedded moonlets inducing these S-shaped structures are not massive
enough to counteract the viscous ring diffusion to open and maintain circumferential
gaps, distinguishing them from ring-moons like Pan and Daphnis.
The density structures caused by ring-embedded moonlets were first studied in
\citet{spahn2000aaa} and \citet{sremcevic2002mnras}. They predicted two radially
shifted gaps with limited azimuthal extent -- the partial propeller gaps. In
addition, they derived scaling laws stating that the radial propeller dimensions
scale with the Hill radius
\begin{equation}
\rhill = a \left( \frac{M_m}{3 M_p} \right)^{1/3}
\label{eq:hill_radius}
\end{equation}
of the respective moonlets and the azimuthal extent of the partial propeller gaps
scales with $\rhill^3/\nu_0$, where $a$ and $M_m$ denote the semimajor axis and the
mass of the propeller-moonlet, whereas $M_p$ labels the mass of Saturn and $\nu_0$
the kinematic shear viscosity of the ring.
N-body box simulations completed the \emph{fingerprint} of propeller structures
by revealing moonlet wakes adjacent to the partial propeller gaps \citep{seiss2005grl,
sremcevic2007nature, lewis2009icarus}.

Recent stellar occultation scans of Cassini's Ultraviolet Imaging Spectrometer (UVIS)
were able to resolve both features for the largest propeller structure named \bleriot{}
\citep{seiss2018aj, baillie2013aj}. \bleriot{} is by far the most imaged trans-Encke
propeller and until recently only \bleriot{} was known to show well-formed partial gaps
in a few high-resolution images taken by Cassini's Narrow Angle Camera (NAC)%
\footnote{Since then, partial gaps were also resolved for the trans-Encke propellers
Earhart and Santos-Dumont in very high-resolution observations taken during Cassini's
Ring Grazing Orbits.}, the analysis of which will be the focus of this paper.

%
% Bleriots Halbachse (Wert)
%

% What are propellers and where to find them? + One of Cassini's biggest successes ...
% Where do we find propellers (propeller belts, trans-Encke propellers, maybe peggy, maybe B-ring propellers)
% Morphology of a propeller, in particular propeller gaps!
% Some of the trans-Encke propellers are trackable
% which observations are there? (Bleriot, Santos-Dumont, Earhart), very high-resolution, but also slightly lower resolution of Bleriot
%Length of propeller gaps (example Bleriot)
% Table with the observations used in this paper (col: lit/unlit side)
% for funding proposal: picture of Bleriot and plot of azimuthal evolution showing 3200 km long gap

\section{Radial gap profiles from Cassini ISS images}
\label{sec:radial_profiles}

We analyze images of the sunlit side of Saturn's outer A ring in which the
propeller gaps of \bleriot{} are clearly visible%
\footnote{Calibrated images taken by the Cassini space probe can be downloaded
from the Planetary Ring-Moon Systems Node \url{https://tools.pds-rings.seti.org/opus}.}.
We use a coordinate frame centered on the propeller moonlet, denoting the radial
distance to the moonlet with $x$ and the azimuthal distance with $y$.

To calculate radial profiles for a given azimuthal location, we radially bin the
ring with bin-width $\delta x$ and azimuthally average the I/F values over the
azimuthal range $\delta y$ to reduce noise. To calculate the I/F value for a
given bin, the I/F values of the pixels are weighted according to the area the
respective pixels share with the bin.

\begin{figure*}[t]
\centerline{\makebox[0pt][c]{
   \includegraphics[width=14.6cm]{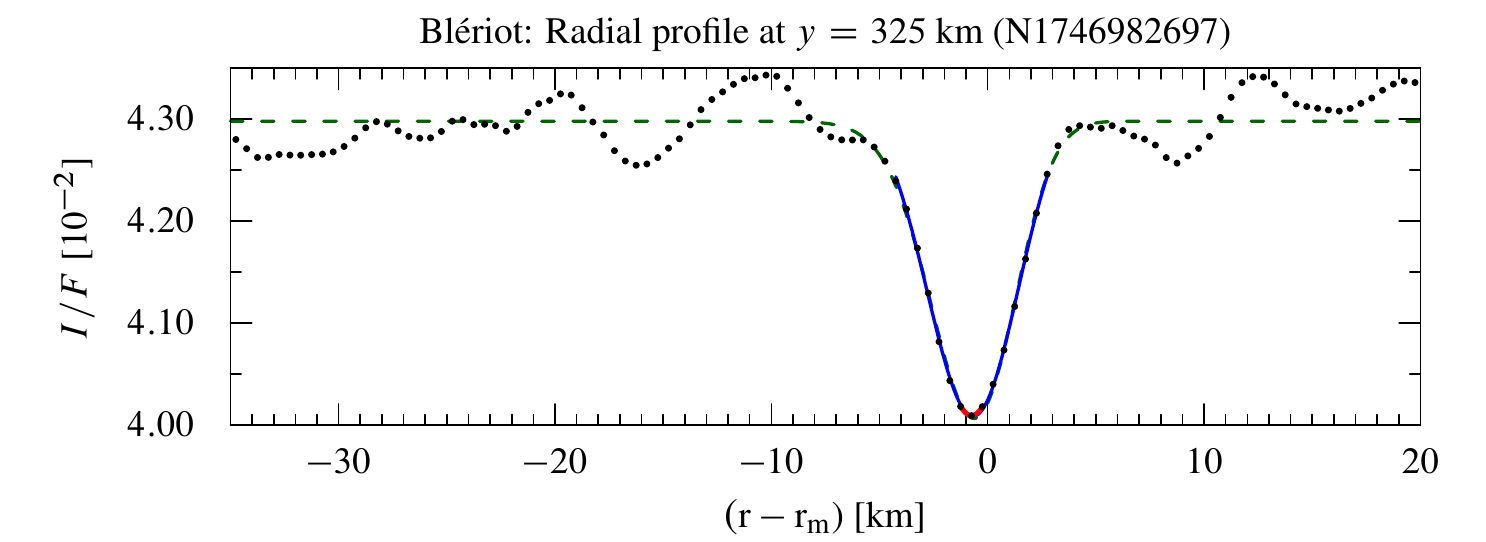}
}}
\caption{\small%
Radial profile of the propeller \bleriot{} 325 km downstream of the moonlet, with
clearly visible propeller gap. In order to determine the radial position of the
gap-minimum and the gap width, we fit different functions to the radial profile.
Shown are fits of \cref{eq:heuristic_fit} to the complete radial range shown in the
plot (dashed dark green curve) and to only the radial range of the propeller gap
(blue solid curve). Additionally, the red solid curve is a parabola determined
from the data point with the lowest I/F value and its two neighbouring data points.
}
\label{fig:radial_profile}
\end{figure*}

\cref{fig:radial_profile} shows a radial profile of \bleriot{} 325 km downstream
of the moonlet. In order to determine the radial location of the gap minimum in a
radial profile, we combine two approaches. First, we determine a parabola from the
data point with the lowest I/F value in the gap region and its two neighbouring
data points. This \emph{local} approach is illustrated by the red curve in
\cref{fig:radial_profile}.

Additionally, we fit the function
\begin{equation}
I/F\left(\xi(x)\right) = \left\{
\begin{array}{cc}
A_0 - \Delta A \exp\left( - \frac{\left[\ln(\xi)\right]^2}{2\eta^2\sigma_\text{gap}^2} \right) & \quad\xi > 0 \\
A_0 & \quad\xi \le 0
\end{array}
\right.
\label{eq:heuristic_fit}
\end{equation}
with $\xi(x) = \eta \left(x - x_\text{gap}\right) + 1$, to the radial profile data.
Here, $A_0$ estimates the brightness of the unperturbed ring, $A_0 - \Delta A$ the
brightness at the gap minimum, $x_\text{gap}$ the radial location of the gap minimum,
and $\sigma_\text{gap}$ the width of the gap. The parameter $\eta$ controls the asymmetry
around the gap minimum. For $\eta\to 0^+$ \cref{eq:heuristic_fit} turns into a Gaussian
profile.

This approach, illustrated by the green dashed and the solid blue curves in
\cref{fig:radial_profile}, takes the shape of the gap into account.
The dashed green curve exemplifies a fit of \cref{eq:heuristic_fit} to the
complete radial range of the radial profile, whereas the solid blue curve shows
a fit of \cref{eq:heuristic_fit} to the radial range of the propeller gap.

The estimates of the location of the gap minimum as calculated from the parabola,
the fit of \cref{eq:heuristic_fit} over the limited radial range, and the fit of
\cref{eq:heuristic_fit} over the complete radial range are $-746\,\text{m}$,
$-631\,\text{m}$, and $-678\,\text{m}$, yielding a mean value of $\bar{x}_\text{gap} =
-685\,\text{m}$ with a root mean square value of about $60\,\text{m}$. For comparison,
the projected radial ring plane resolution of the image is $3\,\text{km}/\text{pixel}$,
the radial bin-width is $\delta x = 500\,\text{m}$ and the azimuthal averaging range
is $\delta y = 50\,\text{km}$. Depending on the orientation of the propeller in
the image, it is often possible to determine radial profiles with sub-pixel
resolution.

\section{Radial separation of propeller gaps}
\label{sec:radial_gap_separation}

\begin{figure*}[t]
\centerline{\makebox[0pt][c]{
   \includegraphics[width=14.6cm]{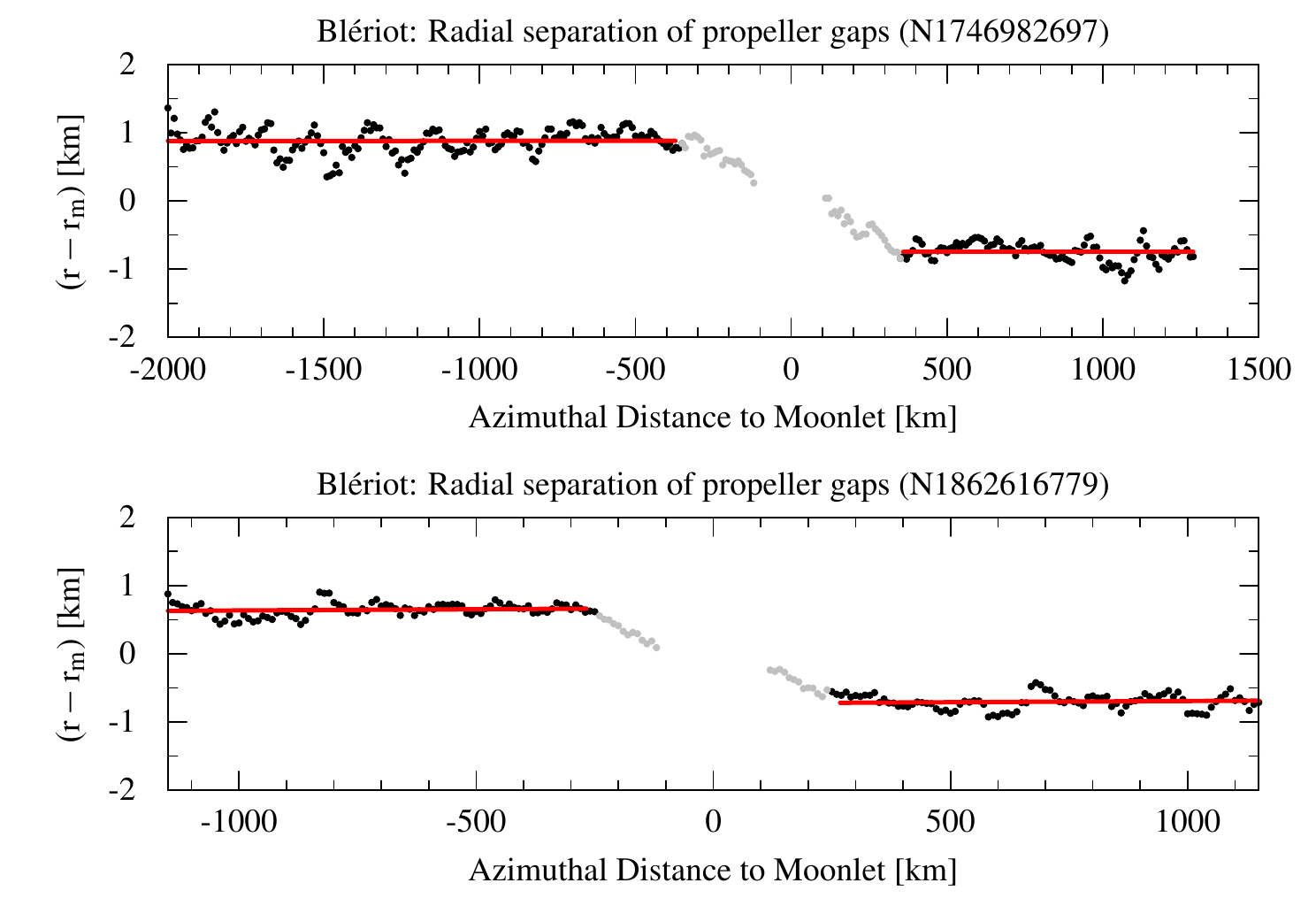}
}}
\caption{\small%
Radial separation of \bleriot's propeller gaps. The plots show the radial distance
of gap minima to the moonlet at different azimuthal locations. The red lines are fits
of \cref{eq:fit_function_radial_gap_separation} to the radial gap minima positions
quantifying the radial separation of the propeller gaps. Only black data points are
used for the fits because the radial gap separation still builds up over the first
few hundred kilometers downstream the moonlet.
}
\label{fig:radial_gap_separation}
\end{figure*}

N-body box simulations \citep{seiss2005grl, lewis2009icarus} and hydrodynamical
propeller simulations \citep{seiss2018aj} show that the radial separation $\Delta r$
of the gap-minima of the partial propeller gaps is about 4 Hill radii.

We determine the radial separation of the propeller gaps by fitting the function
\begin{equation}
f(y) = \left\{
\begin{array}{ll}
\alpha y + b + \Delta r & \quad y < 0 \\
\alpha y + b & \quad y > 0
\end{array}
\right.
\label{eq:fit_function_radial_gap_separation}
\end{equation}
to radial gap-minima positions at different azimuthal locations. The gap-minima
positions are obtained from fits to the different radial profiles prepared for
each image as described in Section \ref{sec:radial_profiles}.

% Beschreibung von Fig. 1
\cref{fig:radial_gap_separation} shows two examples of the azimuthal variation
of radial gap-minima positions. For the first few hundred kilometers downstream
the moonlet the radial separation is still building up (gray data points).
Afterwards, the gap-minima positions fluctuate around a mean value (black data
points). The red lines illustrate fits of \cref{eq:fit_function_radial_gap_separation}
to the black data points yielding $\Delta r$.

% Ergebnisse und Vergleich mit anderen Vorhersagen
Table \ref{table:hill_radius_results} lists the fit results for images taken by
Cassini's narrow angle camera (NAC) which show \bleriot{} with clearly visible
propeller gaps. % Both gaps have to be visible for several hundred kilometers.
These values lead to a Hill radius estimate of $\rhill = (400 \pm 100)\,\text{m}$
for \bleriot.
\citet{seiss2018aj} matched results of hydrodynamical propeller simulations to
two stellar occultation scans performed by Cassini's UVIS instrument that scanned
through the propeller \bleriot. They determined a Hill radius of $\rhill = (600
\pm 100)\,\text{m}$ which is larger than our value but lies within a factor of two
to each other.

One reason for the difference might be that the relation $\Delta r = 4\rhill$ is
only approximately valid and that the radial gap-minima separation depends slightly
on the azimuthal distance to the moonlet as hydrodynamical simulations suggest.
The solution could be to carry out the same analysis performed for the image data
for results of these hydrodynamic propeller simulations, but this is still work in
progress.

\begin{deluxetable}{lcc}
%\tabletypesize{font size command}
\tablewidth{14cm}
\tablecaption{Results of radial gap separation fits for \bleriot\label{table:hill_radius_results}}
\tablehead{
   \colhead{Image} &
   \colhead{\parbox[t]{4cm}{\centering Radial separation\\ $\Delta r$ [km]}} &
   \colhead{\parbox[t]{4cm}{\centering Hill radius\\ \rhill{} [m]\strut}} 
}
\startdata
N1544842586 & 1.47    & 370 \\
N1586641169/1255 & 1.95 & 510 \\
N1731354160 & 1.78    & 445 \\
N1731354280 & 1.65    & 410 \\
N1746982697 & 1.63    & 410 \\
N1862616735 & 1.39    & 350 \\
N1862616779 & 1.40    & 350 \\
\enddata
\end{deluxetable}

\section{Azimuthal evolution of propeller gaps}

\subsection{Analytic gap evolution model}
\label{sec:analytic_gap_evolution}

Let $\Sigma$ be the ring's surface mass density. We will assume a power law
dependence of the kinematic shear viscosity $\nu$ on the surface mass density
\begin{equation}
\nu(\Sigma) = \nu_0 \left(\frac{\Sigma}{\Sigma_0}\right)^{\!\beta}\ ,
\label{eq:shear_viscosity}
\end{equation}
where $\nu_0$ and $\Sigma_0$ are the shear viscosity and surface mass density
of the unperturbed ring. The azimuthal gap relaxation downstream of the propeller
moonlet can be described by a linearized viscous diffusion equation \citep{sremcevic2002mnras}
\begin{equation}
\frac{\partial\!\:\Sigma}{\partial (y/\ak)} = 
   - (x/\rhill)^{-1} \frac{\partial^{\!\:2}\!\:\Sigma}{\partial (x/\rhill)^2}\ 
\label{eq:linear_diffusion}
\end{equation}
where \ak{} defines the characteristic azimuthal length scale of the viscous
diffusion 
\begin{equation}
\ak = \frac{\Omega_0 \rhill^3}{2 (1+\beta) \nu_0}\ .
\label{eq:azimuthal_scaling}
\end{equation}
This scaling has been confirmed by N-body \citep{seiss2005grl} and hydrodynamical
simulations \citep{seiss2018aj} of propeller structures.

\citet{sremcevic2002mnras} derived an approximate Green's solution which solves
\cref{eq:linear_diffusion} for $\sigma(x, y=0) = \delta(x - x_0)$ (where $\sigma
= \Sigma - \Sigma_0$), which reads
\begin{multline}
G(x, y; x_0) = - \frac{\sqrt{3} x_0}{2\rhill}
                 \left( \frac{3y}{\ak} \right)^{-2/3} 
                 \exp \left[ \frac{(x/\rhill)^3 + (x_0/\rhill)^3}{9 (y/\ak)} \right]
                 \text{Bi} \left[ \left(\frac{3y}{\ak}\right)^{-2/3} \frac{x_0 x}{\rhill^2} \right]\ .
\end{multline}
Here, $\text{Bi}$ denotes the Airy function and $\sigma(x, y=0)$ models the
scattering of the ring particles by the propeller moonlet.

The general solution is then given by
\begin{equation}
\Sigma(x, y) = \Sigma_0 + \int \sigma(x_0, y=0)\,G(x, y; x_0)\,dx_0\ ,
\label{eq:mass_density_solution}
\end{equation}
where we will use $\sigma(x, y=0)$ determined from test particle integrations of
the Hill problem \citep{sremcevic2002mnras, hoffmann2015icarus}.

\subsection{Analysis of Cassini NAC images}

\begin{figure*}[t]
\centerline{\makebox[0pt][c]{
   \includegraphics[width=14.6cm]{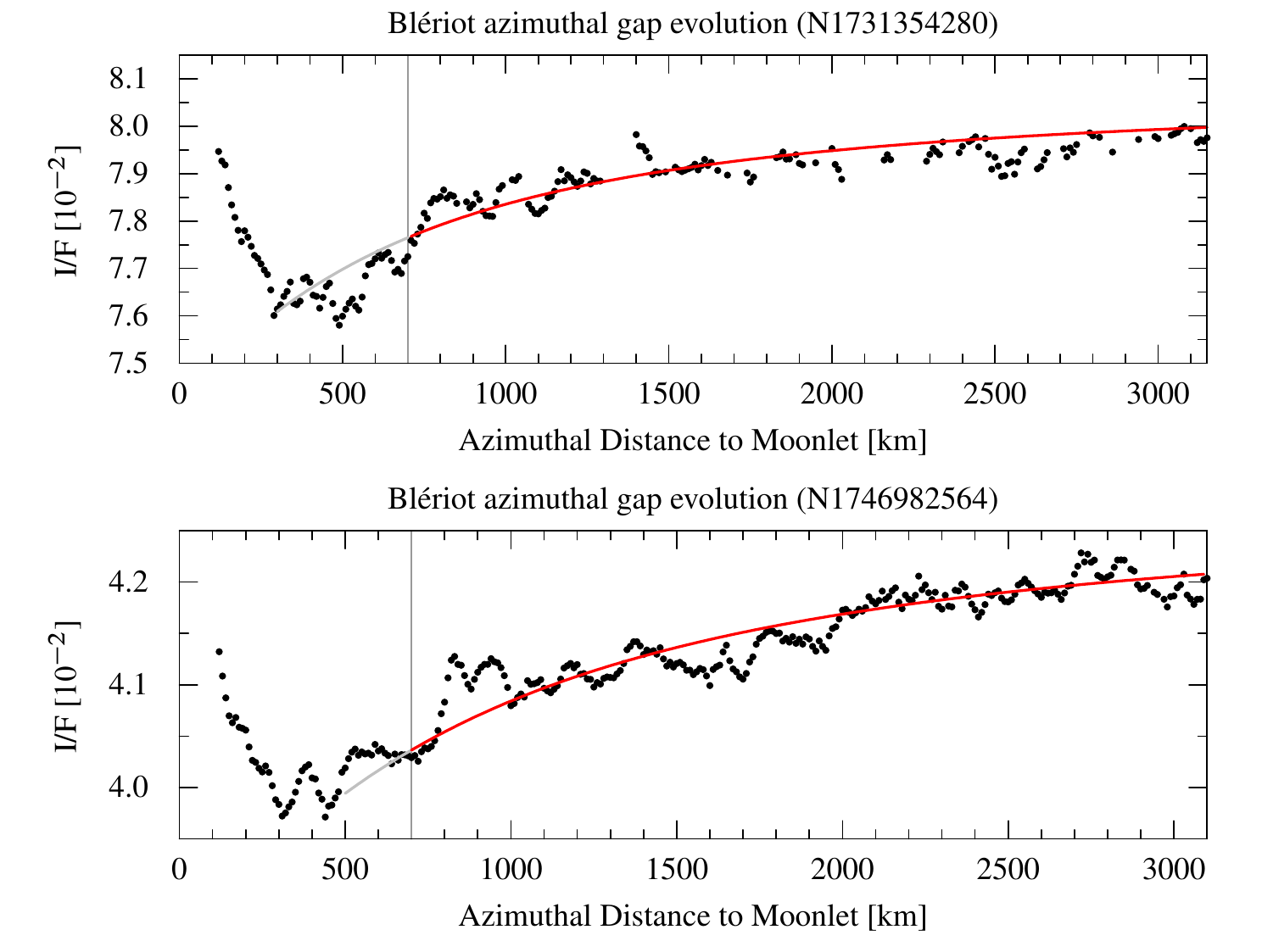}
}}
\caption{\small%
Azimuthal evolution of \bleriot's propeller gaps. The plots show the minimal
brightness values in the propeller gap at different azimuthal locations,
exemplifying that \bleriot's gaps azimuthally extend for over 3000\,km. The
red curves illustrate fits of the surface mass density (\cref{eq:mass_density_solution})
to the brightness values, where we used a single-scattering solution to relate
the ring's surface mass density to I/F values. The vertical gray lines mark
the starting points of the fits ensuring enough azimuthal distance to the
vertically excited region of the propeller, which would not be well described
by the isothermal model \cref{eq:mass_density_solution} is based on.
}
\label{fig:gap_evolution}
\end{figure*}

We determine azimuthal gap profiles by finding the minimal brightness value in
the propeller gap for different azimuthal locations from radial profiles.
\cref{fig:gap_evolution} shows two examples of the azimuthal evolution of \bleriot's
propeller gaps extending for over 3000 km.\footnote{The gaps azimuthally extent
further than shown in \cref{fig:gap_evolution}, the azimuthal range shown limited
by the image border.} The red curves illustrate fits of the analytical model
(\cref{eq:mass_density_solution}) to the azimuthal brightness profiles.

Because we limit our analysis to images of the lit side of the ring, we use a single
scattering solution to relate the relative brightness $I/I_0$ of the azimuthal
profiles to the surface mass density $\Sigma$ of the analytical model
\begin{equation}
\frac{I}{I_0} = \frac{1 - \exp\left[-\tau\,\frac{\mu+\mu_0}{\mu\mu_0}\right]}
                     {1 - \exp\left[-\tau_0\,\frac{\mu+\mu_0}{\mu\mu_0}\right]}\ ,
\label{eq:single_scattering}
\end{equation}
where $\mu = |\sin(B)|$ and $\mu_0 = |\sin(B_0)|$, $B$ and $B_0$ being the
elevation angles of the observer and the sun, respectively. Further, we assume
that the optical depth $\tau$ is proportional to the surface mass density $\Sigma$,
so that $\tau = \tau_0\,(\Sigma/\Sigma_0)$. % How do we determine $\tau_0$?

In \cref{fig:gap_evolution}, the fits start $700\,\text{km}$ downstream of the
moonlet (depicted by the vertical gray lines) in order to be far away from the
vertically excited region of the propeller \citep{hoffmann2015icarus, hoffmann2013apjl},
which would not be well described by the isothermal model \cref{eq:mass_density_solution}
is based on.

The fits result in estimates of the characteristic azimuthal length scale \ak{}
which can be used to predict the ring's kinematic shear viscosity 
\begin{equation}
\nu_0 = \frac{\Omega_0 \rhill^3}{2 (1+\beta) \ak}\ ,
\end{equation}
provided values of the Hill radius $\rhill$ and the power law exponent $\beta$
from \cref{eq:shear_viscosity} are known.
Typical values for $\beta$ determined from N-body box simulations range from
$\beta = 0.67$ for a ring without self-gravity wakes to $\beta = 2$ for fully
developed self-gravity wakes \citep{daisaka2001icarus, salo2018cup}.
Further, we use the in Section \ref{sec:radial_gap_separation} estimated Hill
radius of $400\,\text{m}$ to calculate the kinematic shear viscosity of the
ring in  the vicinity of \bleriot.

Table \ref{table:viscosity_results} summarizes estimates for the propeller \bleriot.
From these values we determine $\ak = (235 \pm 60)\,\text{km}$ leading to
$\nu_{\beta=0.67} = (100 \pm 30)\,\text{cm}^2/s$ and $\nu_{\beta=2} = (60 \pm 20)\,\text{cm}^2/s$.
These estimates are consistent with the viscosity parametrization used by
\citet{daisaka2001icarus}. 
 
\bleriot{} orbits between the Encke and Keeler gaps, a region for which viscosity
estimates are still sparse. For the Encke gap edge \citet{tajeddine2017apjs} estimate
$64\,\text{cm}^2/s$ and \citet{graetz2018apj} find $127\,\text{cm}^2/s \le \nu_0 \le
78\,\text{cm}^2/s$ for $0 \le \beta \le 2$. These estimates agree very well with
the values we determined. On the other hand, estimates for the Keeler gap edge,
$14\, \text{cm}^2/s$ \citep{tajeddine2017apjs} and $32\,\text{cm}^2/s \le \nu_0
\le 20\,\text{cm}^2/s$ \citep{graetz2018apj} are smaller than our values.

\begin{deluxetable}{lccc}
%\tabletypesize{font size command}
\tablewidth{14cm}
\tablecaption{Results of gap evolution fits for \bleriot\label{table:viscosity_results}}
\tablehead{
   \colhead{Image} &
   \colhead{$\ak$ [km]} &
   \colhead{$\nu_{\beta=0.67}\,[\text{cm}^2/\text{s}]$} &
   \colhead{$\nu_{\beta=2}\,[\text{cm}^2/\text{s}]$}
}
\startdata
N1731354160 & 216    & 110 & 61\\
N1731354280 & 224    & 106 & 59\\
N1746982564 & 275    & \phn87 & 48\\
N1746982697 & 276    & \phn86 & 48\\
N1862616735 & 198    & 120 & 67\\
N1862616823 & 222    & 107 & 60\\
\enddata
\end{deluxetable}

\citet{seiss2018aj} find a value of $\nu_0 = (340 \pm 120)\,\text{cm}^2/s$ by
matching isothermal hydrodynamic propeller simulations to UVIS stellar occultation
scans.
This significantly higher value is very likely overestimated because of the
simplified isothermal description of the ring (the UVIS scans intersect the
propeller in the vertically excited region of the propeller).

% viscosity is proportional to \rhill^3 !!!

\section{Conclusions}
% Comparison of Hill radius with: (i) Martin Hydro Paper, (ii) Miodrag Nature 2007 + Tiscareno 2010

The main results of this work are:

\begin{enumerate}
%
%\item Propeller gaps
%
\item From the radial separation of \bleriot's propeller gaps in images taken
   by the Narrow Angle Camera onboard the Cassini spacecraft we estimate the
   Hill radius of \bleriot's propeller moonlet to be $(400 \pm 100)\,\text{m}$.
\item By fitting the analytic solution from \citet{sremcevic2002mnras}, which
   describes the azimuthal evolution of the surface mass density in the propeller
   gap region, to the azimuthal evolution of the propeller gap obtained from
   Cassini NAC images, we estimate the kinematic shear viscosity in \bleriot's
   ring region to be between $\nu_{\beta=2} = (60 \pm 20)\,\text{cm}^2/s$ and
   $\nu_{\beta=0.67} = (100 \pm 30)\,\text{cm}^2/s$.
\end{enumerate}

% The Hill radius we estimate for \bleriot{} is smaller than the value determined
% by \citet{seiss2018aj}. One reason for the difference might be that the relation
% $\Delta r = 4\rhill$ is only approximately valid and that the radial gap-minima
% separation depends slightly on the azimuthal distance to the moonlet. The solution
% could be to carry out the same analysis performed for the image data for results
% of hydrodynamic propeller simulations, which we leave for future work.

\section*{Acknowledgements}
\vspace{-0.5cm}
\acknowledgements We kindly acknowledge the efforts of the Cassini ISS team in
the design and operation of the ISS instrument. This work was supported by the
Deutsche Forschungsgemeinschaft (Ho5720/1-1, Sp384/28-1) and the Deutsches
Zentrum für Luft- und Raumfahrt (OH 1401).

\end{document}